\documentclass[Sensors,article,accept,moreauthors,pdftex]{mdpi} 

\usepackage{algorithm}
\usepackage{algpseudocode}  
\usepackage{caption}

\usepackage{textcomp}
\usepackage{subcaption} 
\def\BibTeX{{\rm B\kern-.05em{\sc i\kern-.025em b}\kern-.08em
    T\kern-.1667em\lower.7ex\hbox{E}\kern-.125emX}}

\firstpage{1} 
\pubvolume{20(5)}

\articlenumber{548}
\doinum{10.3390/s20020548}
\pubyear{2020}
\copyrightyear{2020}
\history{Received: 9 November 2019; Accepted: 13 January 2020; Published: 19 January 2020}

\Title{\textcolor{black}{RDSP: Rapidly Deployable Wireless Ad Hoc System for Post-Disaster Management}
}

\Author{
Ajmal Khan $^{1}$, 
Adnan Munir $^{1}$,
Zeeshan Kaleem $^{2}$,
Farman Ullah $^{1}$,
Muhammad Bilal $^{3}$,
Lewis Nkenyereye $^{4}$,
Shahen Shah $^{1}$,
Long D. Nguyen $^{5}$,
S. M. Riazul Islam $^{6}$
and Kyung-Sup Kwak $^{7,}$*
}

\AuthorNames{Firstname Lastname, Firstname Lastname and Firstname Lastname}

\address{%
$^{1}$ \quad Department of Electrical and Computer Engineering, COMSATS University Islamabad-Attock Campus, 	Attock 43600, Pakistan; drajmal@cuiatk.edu.pk (A.K.); adnanmunir294@gmail.com (A.M.); farmankttk@cuiatk.edu.pk (F.U.); shahenshaah111@gmail.com (S.S.)\\ 
$^{2}$ \quad Department of Electrical Engineering, COMSATS University Islamabad Wah Campus, Wah 47040, Pakistan; zeeshankaleem@gmail.com\\

$^{3}$ \quad Division of Computer and Electronics Systems Engineering, Hankuk University of Foreign Studies, Yongin-si 17035, Korea; m.bilal@ieee.org

$^{4}$ \quad Department of Computer and Information Security, Sejong University, Seoul  05006, South Korea; nkenyele@sejong.ac.kr 

 $^{5}$ \quad Institute of Research and Development, Duy Tan University, Da Nang 550000, Vietnam; nguyendinhlong1@duytan.edu.vn 

$^{6}$ \quad Department of Computer Science and Engineering, Sejong University, Seoul 05006, Korea; riaz@sejong.ac.kr 

$^{7}$ \quad Department of Information and Communication Engineering, Inha University, Incheon 22212, Korea
}

\corres{Corresponding author: kskwak@inha.ac.kr}

\abstract{In post-disaster scenarios, such as after floods, earthquakes, and in war zones, the cellular communication infrastructure may be destroyed or seriously disrupted. In such emergency scenarios,  it becomes very important for first aid responders to communicate with other rescue teams in order to provide feedback to both the central office and the disaster survivors. To address this issue, rapidly deployable systems are required to re-establish connectivity and assist users and first responders in the region of incident. In this work, we describe the design, implementation, and evaluation of a rapidly deployable system for first response applications in post-disaster situations, named RDSP. 
	The proposed system helps early rescue responders and victims by sharing their location information to remotely located servers \textcolor{black}{by utilizing a novel routing scheme. This novel routing scheme consists of the Dynamic ID Assignment (DIA) algorithm and the Minimum Maximum Neighbor (MMN) algorithm. The DIA algorithm is used by relay devices to dynamically select their IDs on the basis of all the available IDs of networks. Whereas, the MMN algorithm is used by the client and relay devices to dynamically select their next neighbor relays for the transmission of messages.} The RDSP contains three devices; the client device sends the victim's location information to the server, the relay device relays information between client and server device, the server device receives messages from the client device to alert the rescue team. We deployed and evaluated our system in the outdoor environment of the university campus. The experimental results show that the RDSP system reduces the message delivery delay and improves the message delivery ratio with lower communication overhead.}

\keyword{disaster management services; device-to-device communication; multi-hop relaying; WiFi; ad hoc network; GPS}

\begin{document}


\section{Introduction}
Cellular telephony is an extensively used communication technology. There are approximately eight billion active cellular subscriptions globally with approximately half of those users added in the last few years, mostly in developing areas \cite{b1}. Currently, mobile-cellular subscribers are more than the total population of the world. This is because people enjoy more than one subscription to take advantage of competing data plans of different cellular operators and so forth. Therefore, in many developing areas, cellular networks have replaced conventional landline telephone systems because of easy usage and the low cost of deployment. However, natural or man-made disasters can disrupt or fully destroy the cellular and land line telecommunication infrastructures and services in the effected areas. Recently, in September 2014, Hurricane Odile struck Mexico's Baja California coasts \cite{b2}. The hurricane of category four destroyed many towns, causing the massive destruction of the electrical infrastructure that left over 90\% of the population without electricity. The destruction of the communication infrastructure resulted in the absence of cooperation between the aid organizations; consequently, thousands of individuals endured hardship, 
 due to mismanaged rescue operations.
 After the distractions of a disaster, the communication among rescue groups, such as firemen, police officers, and paramedics is vital. In particular,  the feedback of first rescue responders is extremely important for an effective rescue and restoration operation. From a networking perspective, the aim is to recover connectivity to offer at least temporary communication services to rescue groups. One way to overcome these issues is to arrange network components, such as relays, access points, or routers to create a temporary network on request \cite{b3}. This needs a quickly deployable network to perform the required relief efforts, including helicopters and first responders on the floor that can save many lives. In addition, to guarantee the safety of survivors in the disaster-affected region, it is very important to manage stranded people's requests in a timely manner to provide a general picture of the total injuries, relocation method, emergency needs, and so forth \cite{b4,b5}. Furthermore, in an attempt to rapidly handle and deliver food and other resources to displaced inhabitants, the need for a secure and reliable communication network that is easy to deploy and relies on radio waves instead of a data cable would be a good choice for communication in disaster situations \cite{b6}.
 
In this work, we propose a rapidly deployable system using multi-hop relays for a post-disaster wireless ad hoc network, named RDSP. 
The RDSP aims to reduce the average waiting times for transmitting the rescue groups and victim's location information towards the control server. In addition, the RDSP scheme enables intermediate relay devices to dynamically select their IDs on the basis of the information provided by their neighbor relays and then each intermediate relay selects the best forwarders towards the control server to minimize end-to-end and round-trip message delivery delays. Finally, the client device is used to transmit victim's information via Wi-Fi towards the control server, which then alerts emergency rescue teams to deliver food and other resources to displaced and stuck survivors. 

The remainder of this paper is organized as follows. Section II summarizes the related work. Section III explains the flow charts and algorithms for the client, relay, and server devices. Section IV describes the performance evaluation and compares the performance of the RDSP with the existing scheme. Finally, Section V provides the concluding remarks.

\section{Related Work}
 Improving the emergency response times in post-disaster situations, where the basic communication infrastructure is completely dismantled, is a critical and challenging task. In such a situation, rapidly deployable networks are needed to enable first responders to interact with disaster survivors, each other, and the control room. These networks operate under  challenging conditions, such as power constraints and establishing a network back haul; further, the network  must be easy to deploy, operate, and maintain. To mobilize the smooth transit of rescue teams for providing emergency services in disaster situations, a number of disaster management schemes or rapid deployment systems have been proposed in the literature that mainly focus on emergency communication in order to connect first responders to the control server.

Authors in \cite{b4,b7,b8} demonstrated the importance of rapidly deployable systems in disastrous scenarios, and provided a survey of the number of schemes that offer adequate post-disaster emergency services. They further discussed the features of the existing systems in the post-disaster environment including their advantages and disadvantages. The authors in \cite{b7} also looked at the networking parameters that are essential in a disaster environment, such as routing overhead, topology management, energy efficiency, and multimedia bandwidth usage. 

The authors in \cite{b5} have developed an energy efficient routing protocol that limits the number of duplicate messages transmission to improve the data delivery ratio and extend the operating time of battery-powered devices. The protocol reduces duplicate messages by finding recurring contacts and generates a routing table that uses recurring contacts to transmit a message to a destination. By finding repeated contacts and creating a routing table that utilizes these repeated contacts to send a message to a destination, the proposed protocol considerably reduces the number of control and operation management messages. Owing to reduction in message transmissions, their protocol improved the overall network energy consumption, while maintaining a high delivery rate.

 In \cite{b9}, the authors have proposed an infrastructure independent device-to-device 
 decentralized network system. Various devices, such as GPS, camera, sensors, and transceivers communicate without centralized coordination. The system works without any base station or access point, as devices communicate with each other when the infrastructure is not available because of an accident or emergency. For the communication within a local vicinity, the devices adopt TDMA 
  to assign a specific slot to each device for communication purpose.

 In \cite{b10}, the authors proposed the concept of rescue base stations (RBS). The 
 RBS is a GSM-compatible solar power drop-in communication system especially designed for disaster scenarios. The proposed system consists of a number of disconnected RBS(es) that provide GSM facilities to a number of registered individuals lying in their coverage range. Each RBS locally stores call/SMS data. Since RBS(es) are disconnected and do not share any network link, individuals with android devices act as information carriers between disconnected RBS(es) to transport the necessary information from one RBS to another.

In \cite{b11}, the authors proposed a system to help military officers in critical situations, such as war conditions or attacks in a gangster area. In military warfare, a robust communication system is required so that the military head can communicate with the soldiers and relay the information easily. A mobile robot is utilized to carry and deploy the nodes at the scene of incident. However, nodes have restricted ranges and can be damaged as the robot moves around snags. The proposed system has communication limitations in situations where no line of sight will be available, such as in urban area.

In \cite{b12}, the authors proposed a robot-assisted scheme, which assists the intermediate nodes between source and destination to relay a message over a long distance. The proposed system deploys robots that utilize mesh technology to create autonomous broadband wireless networks. The actions of robots are controlled by relative signal strength indicators (RSSI). By redistributing the network nodes, it is possible to increase the existing system's throughput. The system is adaptable, self-forming, and self-healing.

Energy in rapidly deployable networks is a major constraint, because the intermediate relays consume a high amount of energy while establishing a connection with other nodes, exchanging information, and routing critical information towards the command center \cite{b13,b14}. Since wireless relays  are battery powered and have a limited power supply, an energy-efficient routing protocol is essential for rapidly deployable networks. 

The authors in \cite{b16} outlined an ad hoc airborne communication scheme using balloons, having excellent line-of-sight, wide transmission range, and low interference. The flying balloons create a multi-hop ad hoc network and get access to internet through an internet gateway placed in disaster hit areas. The rescue workers access internet services by connecting the flying balloons. On each balloon, a Global Positioning System (GPS) receiver is installed to find the balloon's position. 

In work \cite{b17}, the authors proposed a Movable and Deployable Resource Unit (MDRU). The goal of the MDRU is to transport resources to a disaster location and set up the network to provide the necessary communication services after a disaster. A van-type MDRU was deployed  to establish a temporary network in disaster zones. The van carries all necessary equipment required to establish ad hoc communication at the disaster area. The authors in \cite{b18} suggested a model to enhance the MDRU-based network's energy resource utilization and spectrum improvement. The proposed model consists of two stages, named topology formation and transmission division. The former stage configures the gateways of the k spectrum and the later stage divides the transmission from the sender gateways to the MDRU resource unit.

In \cite{b19,b20,b21,b22,b23,b24,b25,b26}, the authors proposed various deployment schemes for disaster response to a building collapse, search and rescue, as well as a resource request through ad hoc networks. In \cite{b19}, the authors proposed a novel approach, called Supporting Urban Preparedness and Emergency Response using Mobile Ad hoc Network (SUPER-MAN). The main objective of that approach was to enable structural engineers and first responders to efficiently disseminate a damaged building status and resource request information towards the control sever with minimum possible interference and delays.  The proposed SUPER-MAN system relies on Radio Frequency Identification (RFID) tags to store assessment information during disaster. A Mobile Ad hoc Network (MANET) of RFID tags is established where Dynamic Source Routing (DSR) is implemented as the routing protocol.

\textcolor{black}{In \cite{b20}, the authors proposed a novel method using a ground penetrating radar (GPR) that automatically detects, locates, and characterizes empty space within disaster rubble. In the proposed method, radargrams are preprocessed to segment the boundaries of empty spaces on the basis of radar signal patterns. To search for uncertainties, 95\% confidence intervals are constructed around the segmented boundaries. The geometric relations of the detected boundaries and their signal characteristics are examined to confirm the existence of free space and to improve detection accuracy. Then, the calibrated velocity of a radar wave and its travel time are used to estimate the location and dimension of empty spaces or voids.}

\textcolor{black}{In \cite{b21}, the authors proposed an emergency resource repository portal (E2RP) system, which is a web-based geo-database service that enables on-site and off-site decision makers to access resource information. The whole E2RP framework incorporates a web collaboration service, radio frequency identification RFID tags, a building blackbox system BBS, a geo-database, and a geographic information system GIS. The E2RP framework provides first responders, including civil engineers, a collaboration medium that enables them to actively respond to disasters. The framework also provides access to critical building information through the BBS. RFID tags are used to store building information which is accessible to first responders through the wireless adhoc network. The GIS is used to locate, collect, and distribute resources to first responders.}

In \cite{b22}, a Geographic Information System (GIS)-based framework is proposed that facilitates equipment allocation during disasters. The proposed framework incorporates three subsystems to assist in information gathering and decision making. First, an application is developed that runs on mobile devices to request on-field resources. Second, a resource repository is deployed with a geospatial database that allows a graphical interface to spatially query resources. Additionally, a GIS is introduced that allows for automatic decision making, such as matching resources and identifying routes for resource distribution. The proposed framework incorporates decision models into the system to assist complex decision-making during equipment delivery.

In ~\cite{b23}, the authors proposed a state-of-the-art technique for collecting data and extracting information to avoid disaster-related injury and post-event damage. A database repository based on GIS, called Extreme Events Database Viewer (EEWV), is being developed to store spatial and temporal data that defines communities before and after disasters. This web platform can store multiple geolocated data formats including photographs and 360{\textdegree} videos. A tool was designed to automatically extract photographs from 360{\textdegree} video data. Extracted images provide a manageable data set to efficiently document the characteristics of buildings and the surrounding environment. The propose system's main objective was to find buildings that were vulnerable to floods and storms. To this end, 1950 buildings were filmed passively with a 360{\textdegree} camera mounted on the vehicle. In order to train a deep learning neural network, these extracted building images were used by the neural network to determine whether a building was elevated or not.

In \cite{b24}, the authors proposed a Reliable Routing Technique (RRT) that ensures reliable data delivery towards the destination device, using mobile devices that are carried by moving people in the incident area. Each mobile device broadcasts the received message towards the destination by maintaining a priority list of probable forwarding candidates. The proposed RRT technique guarantees that the second priority candidate will forward the data packet to the destination device if the first priority candidate is unable to forward the data packet due to its mobility, thus ensuring the reliability of data delivery in the network.

The concept of breadcrumbs is introduced by the authors in \cite{b27,b28,b29,b30}. Breadcrumbs are tiny and cost-efficient relay devices. Their only objective is to relay packets between edge nodes. Thus, in disaster scenarios, rescue team members carry several breadcrumb devices along with a mobile radio to communicate with command center via breadcrumb devices. Rescue team members must regularly drop breadcrumb devices as they explore the disaster area in order to retain end-to-end connectivity with the control server. The breadcrumb relays are dropped to create a static ad hoc network on demand. The command center retains contact with the rescue teams members via relays dropped by the rescue team to enlarge the coverage area. The breadcrumb approach guarantees reliable communication, offers an increased coverage area and eliminates the probability of network partitioning.

Extensive research was carried out to address the problem of the decision of the deployment of breadcrumb. Each proposal describes its own deployment algorithms but they have various common features. Some algorithms monitor the link quality by measuring the signal-to-noise ratio (SNR) \cite{b27}, bandwidth \cite{b28}, or received signal strength indicator (RSSI) \cite{b29}. A threshold is set to trigger a deployment event. When the quality of the link falls below this limit, a fresh relay must be dropped by the user. For instance, a pre-defined threshold is used for all applications \cite{b27} and \cite{b29}, whereas in \cite{b28}, the threshold is set based on the bandwidth requirement of each application. In \cite{b29}, the authors have proposed an ad hoc deployment system to efficiently communicate with victims during disasters. In the proposed system, two types of control information are used for the deployment of relay i.e., relay is deployed either because the link quality degradation is detected by the mobile user, or because an explicit relay deployment request is received by the mobile user from its neighbors. The disaster area could be sufficiently large; therefore, relays support multiple hop transmission to leverage end-to-end communication among each other. The command center is located on the outskirts of the incident region and members of the rescue team begin moving from the command center and move in separate directions into the incident scene. It is assumed that mobile users are connected with each other and with the command center at any time. A mobile rescue team member can drop a relay device if required to maintain connectivity. Each mobile user determines where and when a relay device should be deployed by running an algorithm and alerting the host user through some devices, such as blinking or strong light or sound, when it is necessary to unfold the network. Similarly, in \cite{b27}, the authors proposed an algorithm where relays perform a rapid evaluation of a physical layer to decide on the deployment of the next relay. The relay constantly transmits probe packets to the relays that have been dropped before. A receiving relay responds with an acknowledgment packet if it is within the communication range. Then, through ACK 
reception, the transmitting relay calculates the SNR; if the value of the SNR drops below the threshold level, a new relay is dropped.

The breadcrumb approach provides a suitable solution to extend the coverage area for rescue teams in disaster situations. However, this approach does not offer redeployment possibilities because the relays do not have their own mobility. Indeed, as mentioned in the above proposals, the first responders must drop breadcrumb devices to set up an ad hoc network. However, this is not necessarily the perfect solution. This is because when the first responders join the rescue operation, relay deployment is not their first priority. Hence, they may forget to drop a relay or merely miss the deployment signal. To solve this problem, an automatic breadcrumb dispenser is proposed in \cite{b30}. A Utility Function (UF)-based algorithm is proposed that sets criteria to deploy new breadcrumbs automatically. The UF based algorithm works as follows: the requester broadcasts a help message to initiate the algorithm. After receiving a help message, all the neighbors of requester send their data (number of breadcrumbs) to the requester. Following a predefined timeout, the requester calculates the value of each of its neighbors' utility functions and transmits a drop message to a neighbor with the highest UF value to deploy a new breadcrumb. 

As described above, the existing breadcrumb approaches only focus on the efficient deployment of breadcrumbs to enlarge the coverage area and eliminate the probability of network partitioning. These approaches, however, do not provide effective routing schemes to deliver emergency request messages with minimum latency to the command center. Indeed, these breadcrumb approaches utilize existing routing protocols already developed for mobile ad hoc networks. This is not, however, a perfect solution. This is because breadcrumb devices are utilized as battery-driven intermediate relays, and the current routing schemes can rapidly deplete the battery life of these devices. To efficiently utilize the battery life of breadcrumb devices, the network protocol should be designed in a manner to create the shortest multi-hop path between the command server and the rescue team with a reduced control message overhead and a minimum end-to-end message delivery delay. Furthermore, the current breadcrumb approaches do not provide the command server with any data regarding the location of first responders, which ultimately makes it very difficult to find the victims and rescue team positions in post-disaster situations. To the best of our knowledge, no work exists that both designs a unique routing protocol for breadcrumb devices and incorporates location information of victims to determine their distance from the command server. 

Therefore, the major contribution of this study comprises a rapidly deployable system that both delivers request messages to the command server by utilizing novel routing schemes and also manages the location information of victims to calculate their distance from the command server. \textcolor{black}{The novel routing scheme consists of a Dynamic ID Assignment (DIA) algorithm and a Minimum Maximum Neighbor (MMN) algorithm. The DIA algorithm is used by relay devices to dynamically select their IDs on the basis of all available IDs of networks. Whereas, the MMN algorithm is used by the client and relay devices to dynamically select their next neighbor relays for the transmission of messages.} In addition, we provide details of algorithms performed by the client device, relay device, and server device. Furthermore, extensive real time experiments are preformed to demonstrate how the proposed RDSP scheme reduces the control messages' overhead to deliver the request messages with a minimum end-to-end delay and an increased message delivery ratio in post disaster situations.
 
 \section{System Architecture of the RDSP Scheme}
This section presents the architecture of the RDSP scheme, which aims to reduce the average waiting times for transmitting victim's information towards the server by utilizing the following key features: 
\begin{itemize}
\item Client Device: The client devices are the end point communication devices held by rescue team members and victims. The client device is a WiFi-enabled device that manages to transmit the victim's location information to the server device using relay devices. The client devices are used to establish the communication between rescue teams and server and to send feedback about the emergency situation.
\item Relay Device: Relay devices are randomly deployed at a distance of 90 m from each other to transmit the victim's location information generated by the client device towards the server and then send back acknowledgment information to the client device. Relay devices dynamically connect with each other in a manner to establish the shortest path towards the server.
\item Server Device: The server device continuously listens for the arrival of incoming messages sent by the client device via relay devices in order to alert the rescue teams. Moreover, it sends the control and operational messages and is also responsible for receiving the feedback from rescue teams.
\end{itemize}

\begin{figure}[H]
\centering

    \includegraphics[width=0.9\textwidth]{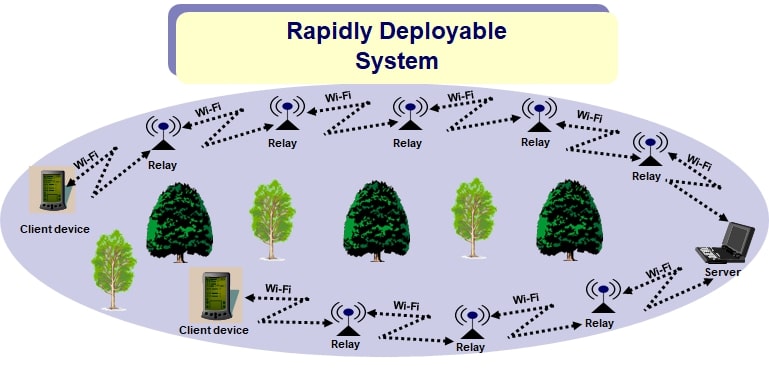}

      \caption{Deployment scenario of the RDSP scheme.}
       \label{fig1}
       
  \end{figure}

Figure \ref{fig1} illustrates a deployment scenario of the RDSP system after disaster. When disaster occurs, the rescue team members will deploy the RDSP system as follows: A server device is installed in the incident area that will receive updates from rescue teams and victims. Additionally, rescue team members will move forward towards disaster areas while deploying relay devices at equal distances of about 90 meters until any victim is sighted. Then, the client device that is carried by the rescue team members is used to send the victim's position information to the server. \textcolor{black}{The distance of 90 meters between relays is managed by incorporating GPS modules in relays. After the server module is installed in an incident area, we start deploying the relays as follows: First of all, the first relay wirelessly connects with the server module. Afterwards, while carrying the relay and moving away from the server module, the relay continuously calculates its distance from the server using the GPS module that provides location coordinates of the relay module, whereas 
 the location coordinates of the server are fixed and known to the relay module. If the calculated distance is 90 meters, a green LED (installed on relay module) lights up, indicating to drop the first relay module at that particular position. Similarly, a second relay is chosen which connects with the first relay and the distance between the first relay and the second relay is again calculated by the second relay and it is dropped when the green LED on the second relay lights up. Following this procedure, all the relay devices manage to maintain a distance of 90 meters between each other.} The 90 meters distance was selected because the relay device uses WiFi technology that has a transmission range of 90 meters. However, the transmission range can be increased by using other advanced technologies, which is highly application dependent. It is indicated in Figure 1 that both the client device and relay device communicate wirelessly with the server device via WiFi. The server device manages all the request messages received and sends back an acknowledgment to the client device to ensure the reception of the request message.

\begin{figure}[H]
  
  \centering
    \includegraphics[width=0.3\textwidth,scale=0.5]{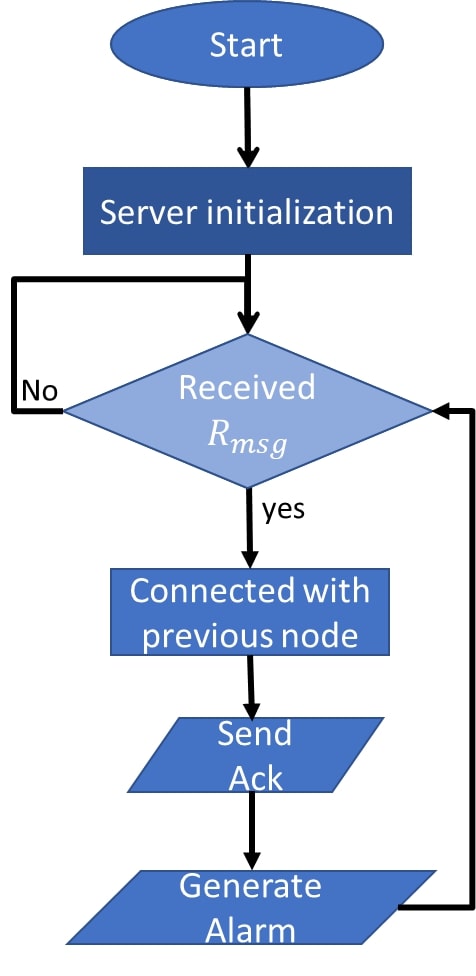}
      \caption{Flow chart of server device.}
       \label{fig2}
  \end{figure}

  Figure \ref{fig2} presents a flow chart of the server device. After the  initialization, the server waits for the arrival of request messages. If a message arrives, the server sends back an acknowledgment message to the client device via relay devices and generates an alarm message to alert the rescue teams.
  
 Algorithm \ref{alg1} presents the operation of  the microcontroller in the server device. The input to the server device is a request message $ R_{msg}$, whereas the output is an acknowledgment message and an alarm message. As shown in step 2, the server device continuously waits for the detection of $ R_{msg}$~. If $ R_{msg}$  is detected, alarm message is generated, acknowledgment is sent back to the client device and $ R_{msg}$  is printed on screen. The microcontroller in the server device
utilizes the Haversine formula \cite{b31} to determine the distance
between the client device and server. This process
is defined by the following Equation (1):
 \begin{align}
&dist=2r{{\sin }^{-1}}\sqrt{  \begin{aligned} {{\sin }^{2}}\left( \frac{\text{ }\!\!\Delta\!\!\text{ }\phi }{2} \right)+\cos \left( {{\phi }_{1}} \right)\text{*}\cos \left( {{\phi }_{2}} \right)\text{*}{{\sin }^{2}}\left( \frac{\text{ }\!\!\Delta\!\!\text{ }\delta }{2} \right)~~~~\end{aligned} }
\end{align} 
where $\phi$ is the latitude, $\delta$ is longitude (in radians).

\begin{algorithm}[H]
\caption{ Server device}

{\bfseries Input:}$ R_{msg}$ (Request message containing node ID, message ID, and GPS coordinates)

{\bfseries Output:} $Ack_{msg}$ (Acknowledgment message )          

 \quad \quad \quad\quad$ Alarm_{msg}$  (Alarm message generated)
\newline \noindent\rule{15.5cm}{0.6pt}

\begin{algorithmic}[1]
\Procedure{}{}   
\smallskip
  \State Step 1: defining and initializing variables
\smallskip
 \State Step 2: detecting events
\smallskip
    \While{1}
    \smallskip
        
        \If{$R_{msg}=true$}
        \smallskip
     
        \State Generate $ Alarm_{msg}$ 
        \smallskip
        \State   send $Ack_{msg}$   back
        \smallskip
        \State  Print request message on screen.
        \smallskip

        \EndIf
        \smallskip
    \EndWhile
    \smallskip

\EndProcedure

\end{algorithmic}
\label{alg1}
\end{algorithm}

\begin{figure}[H]
\centering
    \includegraphics[width=0.66\textwidth]{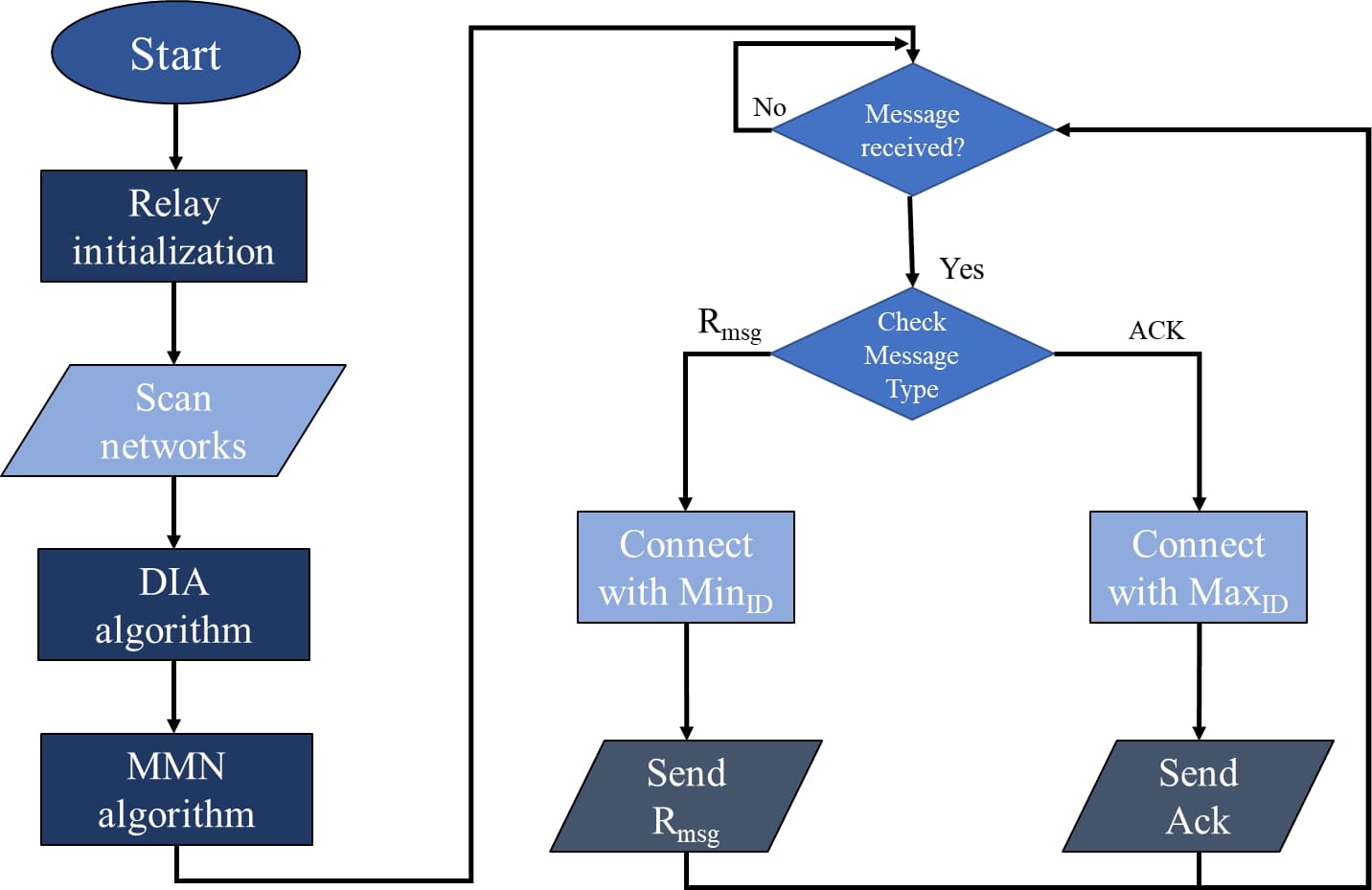}
     \caption{Flow chart of relay device.}
       \label{relay}
  \end{figure}

 Figure  \ref{relay} shows the flow chart of the relay device. After initialization, the relay device scans for available networks and applies the Dynamic ID Assignment (DIA) algorithm (explained later in Algorithm 2) to generate its own ID. Then, the Minimum Maximum Neighbor (MMN) algorithm is utilized that returns minimum and maximum IDs (explained later). Afterwards, the relay device waits for the arrival of messages. If the request message arrives, it selects the minimum ID to transmit the message to the server. The relay with the minimum ID is selected because it is much closer to the server (as will be explained later in the DIA algorithm) and delivers the message in minimum possible time. However, if an Ack 
  message arrives, the relay device selects the maximum ID to transmit the message back to the client. 
 
 \begin{algorithm}[H]
\caption{ Dynamic ID Assignment (DIA)}

{\bfseries Input:} $ AVB_{net}$ (Available Networks)

{\bfseries Output:} $ID_{assigned}$ (Assigned ID )          
 \newline \noindent\rule{15.5cm}{0.6pt}

\begin{algorithmic}[1]
\Procedure{}{} 
\smallskip
  \State Step 1: defining and initializing variables
   \smallskip

      \State      $Array_{id}$  = Array containing available networks IDs
      \smallskip

        \State  $F_{index}$    = First index of $Array_{id}$ 
        \smallskip

         \State   $S_{index}$  = Second index of  $Array_{id}$ 
         \smallskip

 \State Step 2: Assigning ID to relay
 \smallskip

      \State Store $  AVB_{net}$  in  $Array_{id}$
      \smallskip

        \State Sor t$Array_{id}$   in ascending order
        \smallskip

  \State $ID_{assigned}$ = $Array_{id}$ [ $F_{index}$]
\smallskip

        \If{$ID_{assigned}=-1$}
     \smallskip

        \State$ID_{assigned} $= $Array_{id}$ [ $S_{index}$]
\smallskip

        \EndIf
        \smallskip

     \State  $ID_{assigned}$ = $ID_{assigned }$ +1
\smallskip

\EndProcedure

\end{algorithmic}
\label{alg2}
\end{algorithm}
   Algorithm \ref{alg2} presents the dynamic ID assignment (DIA) algorithm. This DIA algorithm is used by relay devices to dynamically select their IDs based on all the available IDs of networks. Initially, it is assumed that the server has an ID = 0 and all the deployed relay devices have an ID = $-$1.  As shown in step 2, each relay device stores available network IDs in $Array_{id}$. Then, $Array_{id}$  is sorted in ascending order. Finally, the relay device generates its own ID by selecting the first index of $Array_{id}$ as it the contains minimum ID. However,  if the first index of  $Array_{id}$  contains -1, then the second index of  $Array_{id}$  is selected and incremented by 1.

\begin{figure}[H]
  \centering
  
    \includegraphics[width=0.65\textwidth]{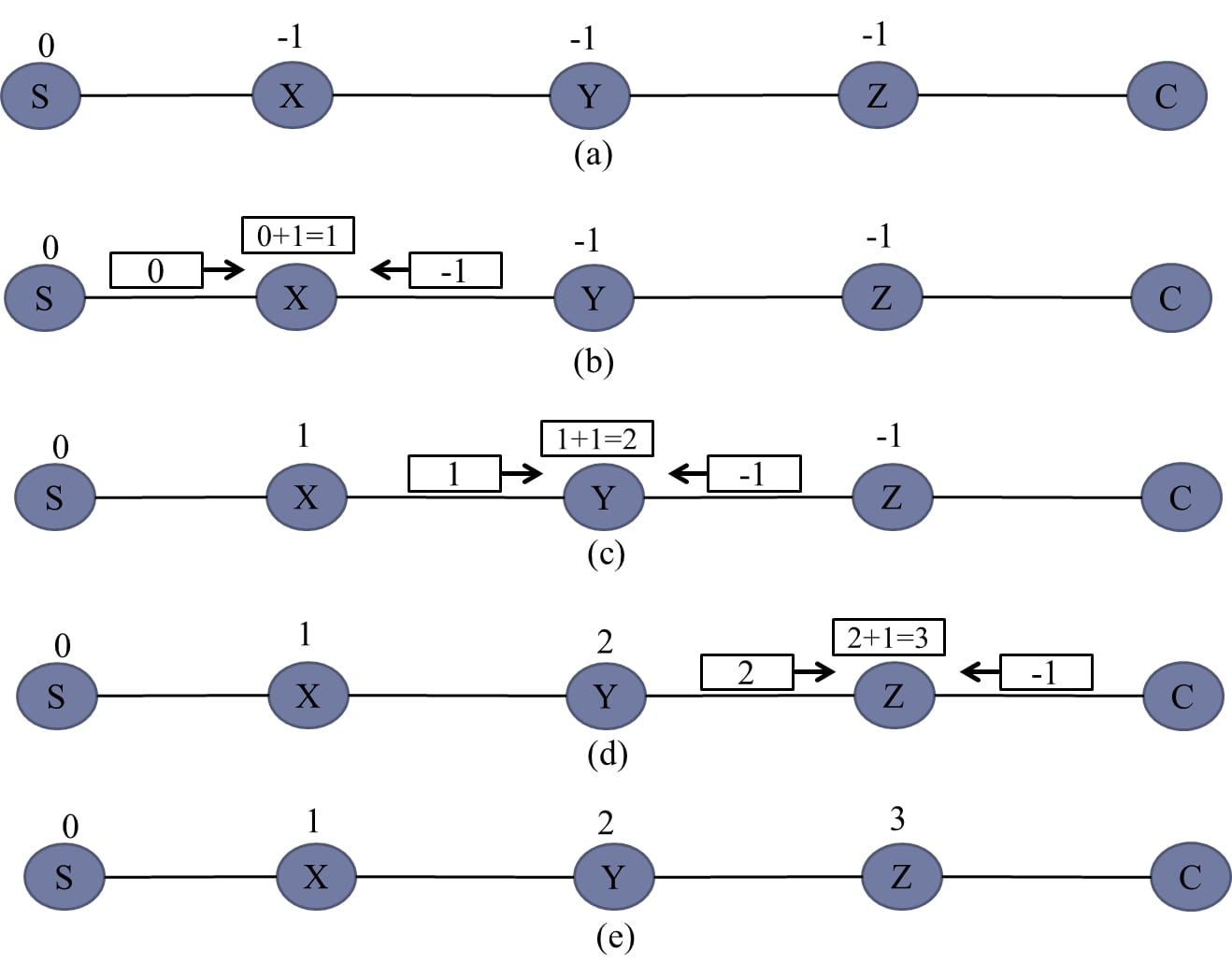}

     \caption{Dynamic ID assignment procedure.}
       \label{figiv}
  \end{figure}

   The detailed procedure of the DIA algorithm is explained in Figure \ref{figiv}. All relays are deployed randomly at a distance of 90m and initially their IDs are $-$1 and the server ID are 0, as shown in Figure \ref{figiv}a. According to Figure \ref{figiv}b, relay X will receive network IDs 0 and $-$1 from server S and relay Y, respectively. By applying the  DIA algorithm, relay X will choose the positive minimum ID, i.e., 0 and increments it by 1.  Therefore, the ID of relay X becomes 1. Similarly, in the same fashion, relay Y will receive IDs 1 and $-$1 from relay X and Z, respectively. As shown in Figure \ref{figiv}c, after applying the DIA algorithm, relay Y will chose the positive minimum ID, i.e., 1 and increments it by 1. Therefore, the ID of relay Y will become 2 and this process will continue until all the relays will be assigned dynamic IDs, as shown in Figure\ref{figiv}e. Hence, it can be seen from Figure \ref{figiv} that relays having lower IDs are much closer to the server as compared with relays having higher IDs.

  Algorithm \ref{algthree} presents the MMN algorithm. This MMN algorithm is used by client and relay devices to dynamically select their next neighbor relays for the transmission of messages. As shown in step 2, each relay device stores available network IDs in $Array_{id}$. Then, $Array_{id}$ is sorted in ascending order. Finally, the relay device finds $ Min_{ID}$  and $Max_{ID}$ by selecting the first and last index of $Array_{id}$, respectively.
  
 \begin{algorithm}[H]
\caption{ Minimum Maximum Neighbor (MMN)}

{\bfseries Input:} $ AVB_{net}$ (Available Networks)

{\bfseries Output:} $Min_{ID}$ and $Max_{ID}$       
 \newline \noindent\rule{8.5cm}{0.6pt}
\algdef{SE}[SUBALG]{Indent}{EndIndent}{}{\algorithmicend\ }%
\algtext*{Indent}
\algtext*{EndIndent}
\begin{algorithmic}[1]
\Procedure{}{}
\smallskip
  \State Step 1: defining and initializing variables
   \smallskip

       \State \hspace*{10mm} $Array_{id}$ = Array containing available networks IDs
       \smallskip
        \State \hspace*{10mm} $F_{index}$    = First index of $Array_{id}$ 
       \smallskip
         \State \hspace*{10mm}  $L_{index}$  = Last index of  $Array_{id}$ 
        \smallskip
     
 \State Step 2: Finding Minimum and Maximum IDs
\smallskip
      \State \hspace*{10mm}Store $  AVB_{net}$  in  $Array_{id}$
       \smallskip
        \State \hspace*{10mm}Sort $Array_{id}$   in ascending order
        \smallskip
  \State \hspace*{10mm} $Min_{ID}$ = $Array_{id}$ [ $F_{index}$]
                \smallskip

        \State \hspace*{10mm} $Max_{ID} $= $Array_{id}$ [ $L_{index}$]
                \smallskip

\EndProcedure

\end{algorithmic}
\label{algthree}
\end{algorithm}

\begin{figure}[H]
  \centering
    \includegraphics[width=0.6\textwidth]{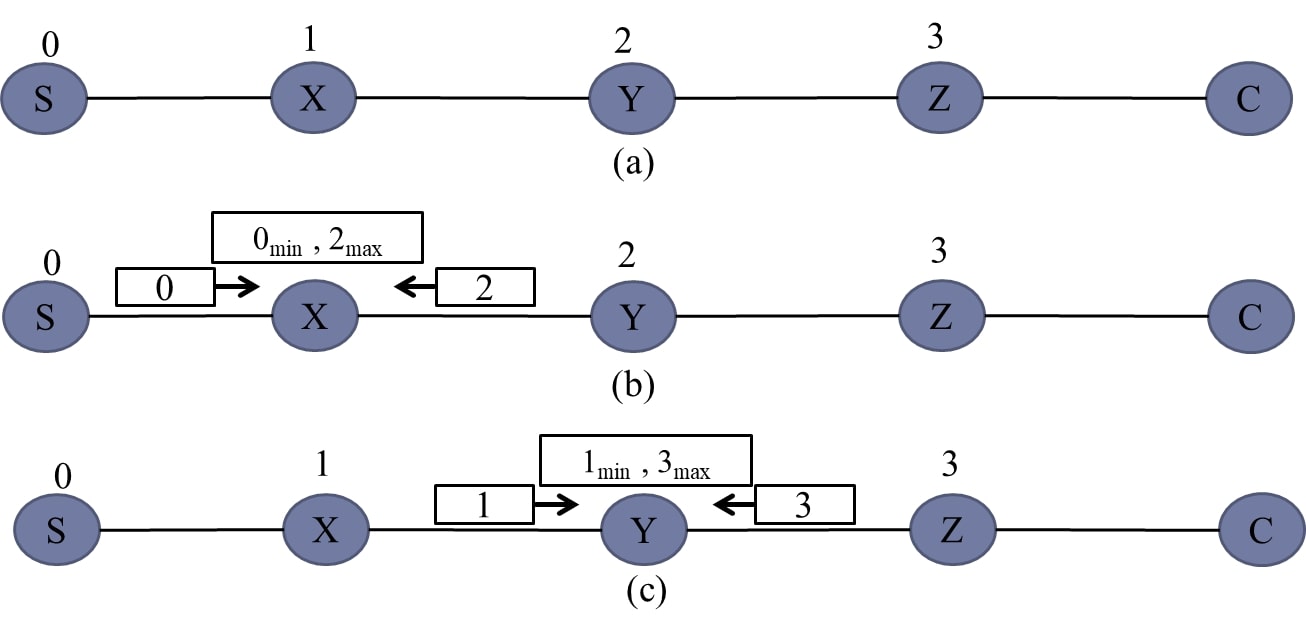}

     \caption{Minimum Maximum Neighbor selection}.
       \label{figv}
  \end{figure}

 The detailed procedure of the MMN algorithm is explained in Figure \ref{figv}. Figure \ref{figv}a shows that all the relays have been assigned IDs after applying the DIA algorithm, as explained earlier in Algorithm~ 2. Afterwards, each relay will select its next neighbor relay for the transmission of messages. It is shown in Figure \ref{figv}b that relay X will receive IDs 0 and 2 from server S and relay Y respectively. By applying MMN algorithm, relay A will choose 0 as the minimum ID and 2 as maximum ID. Similarly, in the same fashion, relay Y will receive IDs 1 and 3 from relay X and Z, respectively. After applying the MMN algorithm, relay Y will chose 1 as the minimum ID and 3 as the maximum ID. 
 
Algorithm \ref{alg4} presents the code for the microcontroller in the  relay device. The input to the relay device is a request message $R_{msg}$, an acknowledgment message  $Ack_{msg}$, and available networks $AVB_{net}$. The  output is either a request message or an acknowledgment message. As shown in step 1,  the relay device scans for available networks and then it applies the DIA algorithm to select its ID and then the MMN algorithm to select minimum and maximum neighbor IDs. Then in step 2, the relay device continuously waits for the detection of events. If $R_{msg}$ is detected,  then the request message is sent to the next relay node having $Min_{ID}$. However, if $Ack_{msg}$ is detected, then acknowledgment is sent back towards the client device via the next relay node having $Max_{ID}$.

\begin{algorithm}[H]
\caption{ Relay device}

{\bfseries Input:}$ R_{msg}$ (Request message containing node ID, message ID and GPS coordinates)

 \quad\quad\quad$Ack_{msg}$ (Acknowledgment message )
  
  \quad\quad\quad$ AVB_{net}$ (Available Networks)

{\bfseries Output:}$ R_{msg}$ (Request message containing node ID, message ID and GPS coordinates)

 \quad\quad\quad\quad$Ack_{msg}$ (Acknowledgment message ) 
\newline \noindent\rule{15.5cm}{0.6pt}

\begin{algorithmic}[1]
\Procedure{}{}  
\smallskip
  \State Step 1: defining and initializing variables
  \smallskip

  \State \hspace*{10mm} Scan $AVB_{net}$
  \smallskip
    \State \hspace*{10mm} Apply DIA Algorithm (assigns ID)
    \smallskip
        \State  \hspace*{10mm} Apply MMN Algorithm ( returns Min \& Max ID)
        \smallskip
 \State Step 2: detecting events
\smallskip
    \While{1}
       \smallskip 
        \If{$R_{msg}=true$}
     \smallskip
        \State send $R_{msg} $ to  $Min_{ID}$
        \smallskip
        \EndIf
        \smallskip
           \If{$ Ack_{msg}=true$}
           \smallskip
     
        \State send $Ack_{msg} $ to  $Max_{ID}$  
        \smallskip
        \EndIf
        \smallskip

    \EndWhile
    \smallskip

\EndProcedure

\end{algorithmic}
\label{alg4}
\end{algorithm}

\begin{figure}[H]
  \centering
  
    \includegraphics[width=0.48\textwidth]{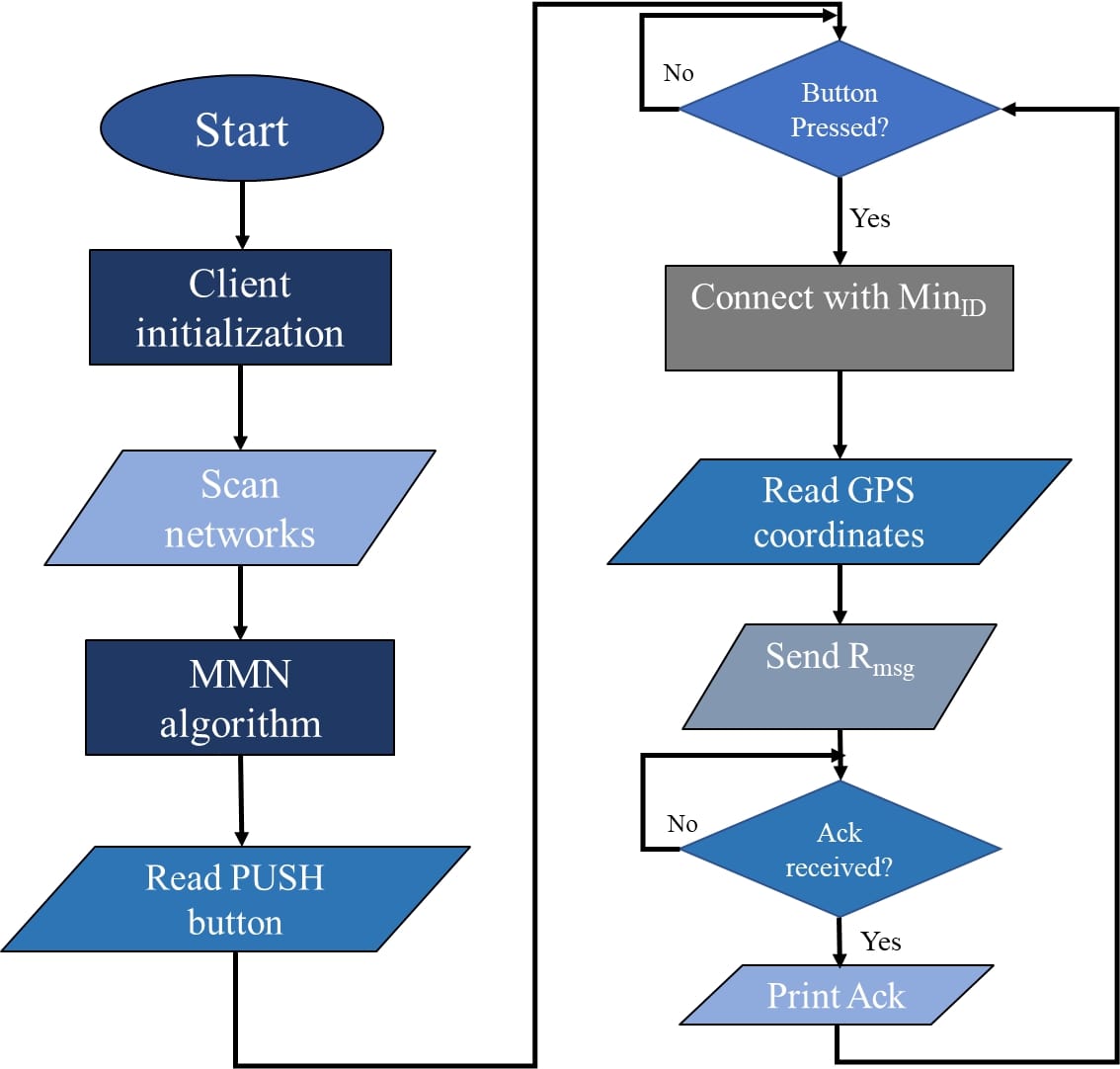}

     \caption{Client device.}
       \label{fig6}
  \end{figure}

Figure \ref{fig6} shows the flow chart of client device. The client device includes a GPS system, a microcontroller, WiFi device, and a push button. The GPS device is utilized to receive the latitude and longitude information of the victim. The microcontroller is utilized to transmit the victim’s position information to the server via the WiFi module. As shown in the flowchart, after initialization, the client device scans for available networks and applies the MMN algorithm that returns the minimum and maximum IDs (explained earlier). Afterwards, the client device reads the status of push button. If the push button is pressed, it connects with the relay node having a minimum ID. The client device then reads the GPS coordinates and transmits the request message towards the server via multi hop intermediate relays. The relay with a minimum ID is selected because it is much closer to the server (as explained earlier in DIA algorithm). After sending the request message, the client device waits for the arrival of the acknowledgment message. If acknowledgment is received, it is then printed on the screen.

Algorithm  \ref {alg5} presents the code for the microcontroller in the client device. The input to the client device is the push button $ P_{b}$, acknowledgment message   $Ack_{msg}$, GPS coordinates $GPS_{cor}$, and available networks   $ AVB_{net}$. The output is a request message $ R_{msg}$. As shown in step 1,  the client device scans for available networks and it applies the MMN algorithm to select minimum and maximum IDs. Then, in step 2, the client device continuously waits for the detection of events. Since, the client device includes a push button that is pressed by a rescue team member if any victim is sighted. Therefore, if the push button $ P_{b}$ is detected,  the client device reads $ GPS_{cor}$ and transmits the request message to the next relay node having $Min_{ID}$  and continuously waits for the arrival of the acknowledgment message. If the $ Ack_{msg}$  is detected, then acknowledgment is printed on the screen.

\begin{algorithm}[H]
\caption{ Client device}

{\bfseries Input:}$ P_{b}$ (Push button)

\quad\quad\quad$Ack_{msg}$ (Acknowledgment message )

\quad\quad \quad$GPS_{cor}$ (GPS coordinates )   

 \quad\quad\quad$ AVB_{net}$ (Available Networks)
 
{\bfseries Output:}$ R_{msg}$ (Request message containing node ID, message ID, and GPS coordinates)
   \newline \noindent\rule{8.5cm}{0.6pt}

\begin{algorithmic}[1]
\Procedure{}{}  
\smallskip
  \State Step 1: defining and initializing variables
  \smallskip
  \State \hspace*{10mm} Scan $AVB_{net}$
  \smallskip
            \State \hspace*{10mm} Apply MMN Algorithm ( returns Min \& Max ID)
            \smallskip
 \State Step 2: detecting events
\smallskip
    \While{1}
        \smallskip
        \If{$P_{b}=true$}
        \smallskip
             \State read  $GPS_{cor}  $
             \smallskip
        \State  send $R_{msg} $ to  $Min_{ID}$  
        \smallskip
        \EndIf
        \smallskip
           \If{$ Ack_{msg}=true$}
           \smallskip
     
        \State Pring $Ack_{msg} $ on screen
        \smallskip
        \EndIf
        \smallskip

    \EndWhile
    \smallskip

\EndProcedure

\end{algorithmic}
\label{alg5}
\end{algorithm}

\begin{figure}[H]
\centering
  \begin{subfigure}[t]{0.7\textwidth}
      \includegraphics[width=0.9\textwidth]{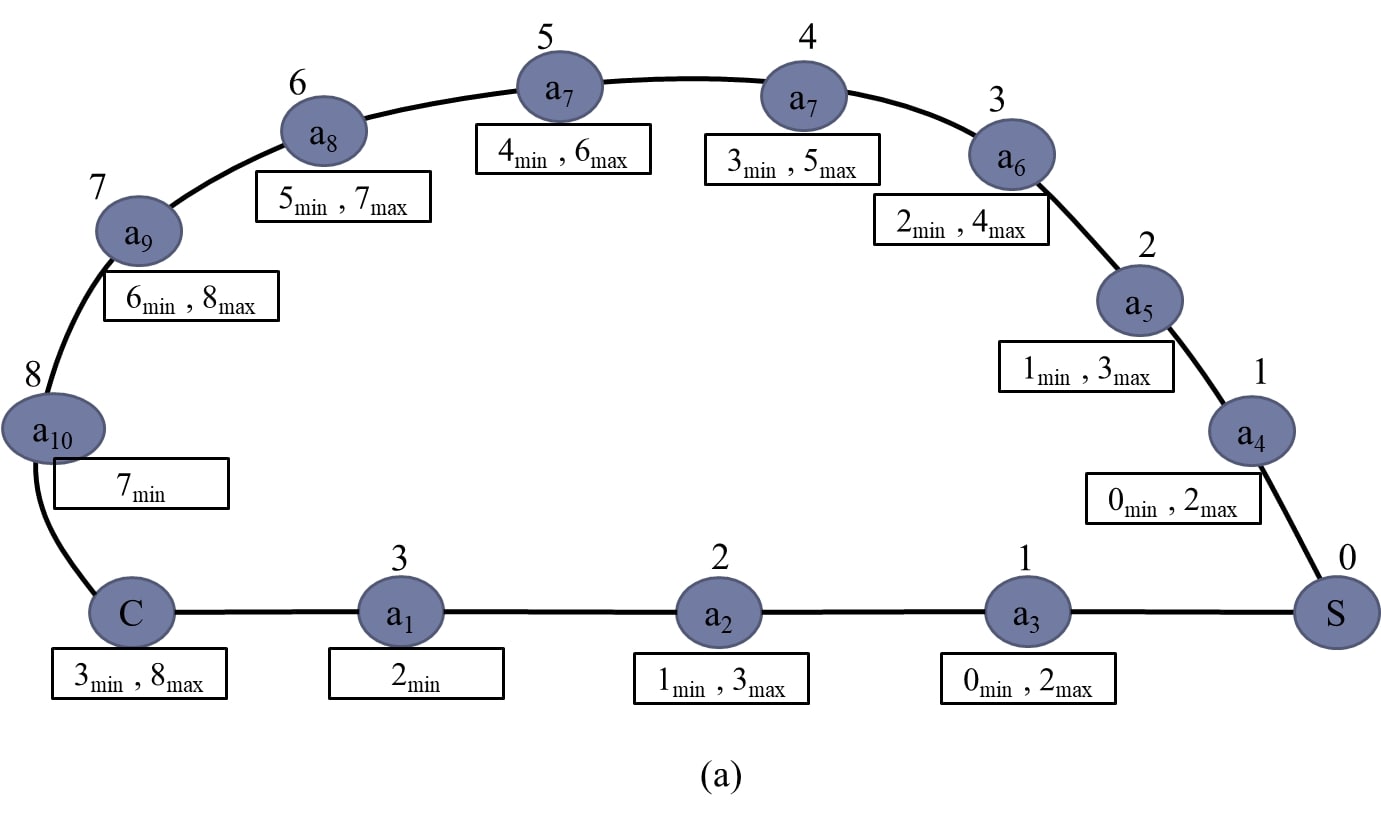}
     \end{subfigure}
   
     \begin{subfigure}[t]{0.7\textwidth}
      \includegraphics[width=0.9\textwidth]{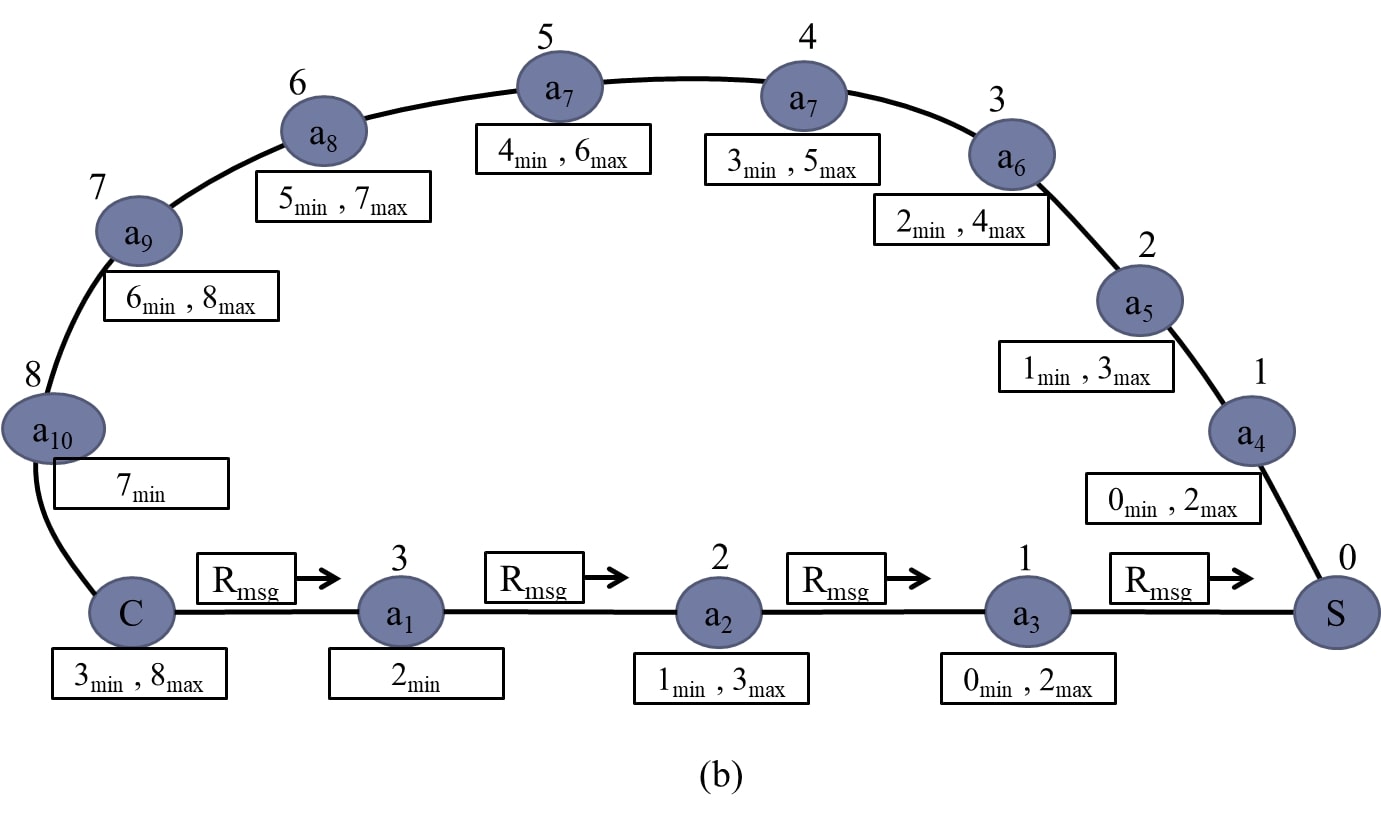}
     \end{subfigure}
    
     \begin{subfigure}[t]{0.7\textwidth}
      \includegraphics[width=0.9\textwidth]{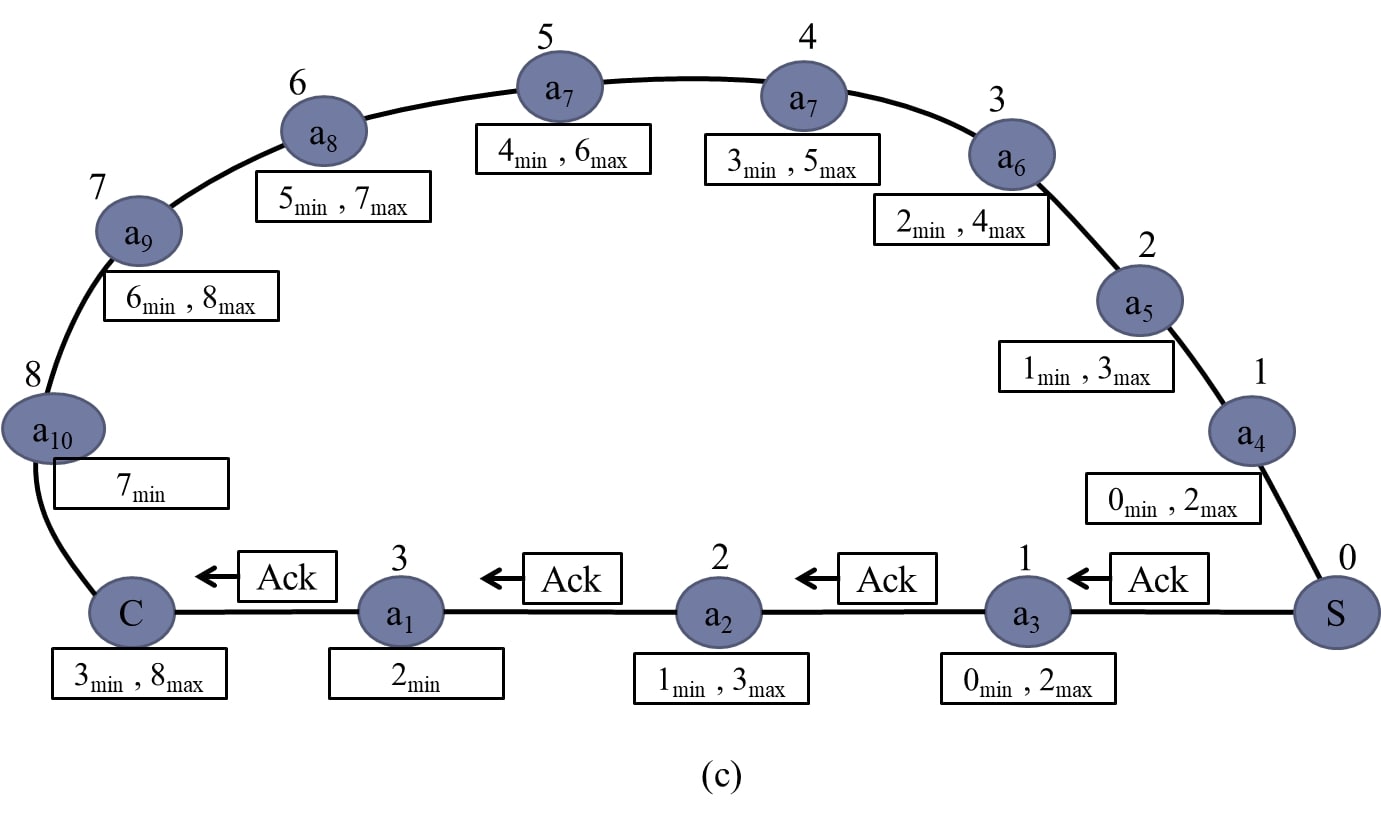}
     \end{subfigure}
    
     \caption{network deployment scenario.}
       \label{fig7}
  \end{figure}

Figure \ref{fig7}a shows a sample network scenario to explain the working procedure of the RDSP system. As shown in the figure, relays are deployed in a manner to establish two different paths between server device S and client device C. The first path is defined by C, $a_1$, $a_2$, $a_3$, S whereas the second path is defined by C, $a_{10}$, $a_9$, $a_8$, $a_7$, $a_6$, $a_5$, $a_4$, S. After deployment, each relay node applies the DIA algorithm to select the dynamic ID based on available networks.

As shown in Figure 7a, after applying the DIA algorithm, the IDs of $a_3$, $a_2$, $a_1$ are 1, 2, and 3 respectively. Similarly IDs of $a_{4}$, $a_5$, $a_6$, $a_7$, $a_8$, $a_9$, $a_{10}$ are 1, 2, 3, 4, 5, 6, 7, and 8 respectively. After ID assignments, each relay and client node applies the MMN algorithm to to select $Min_{ID}$   and  $Max_{ID}$ . Client C receives ID 3 from relay $a_1$ and ID 8 from relay $a_{10}$. Therefore, client C selects 3 as the minimum ID and 8 as the maximum ID. Similarly, relay $a_1$ selects 2 as the minimum ID and relay $a_{10}$ selects 7 as the minimum ID. Both relay $a_1$and relay $a_{10}$ only receive IDs from relay $a_2$ and relay $a_9$ respectively but not from client C as client C has no ID. Relay $a_2$ receives ID 1 and 3 from relays $a_3$ and $a_1$, respectively, and hence, selects 1 as the minimum and 3 as the maximum ID. This process is continued until all relays select their minimum and maximum neighbor IDs.
 
Figure 7b shows how the request message $R_{msg}$ is sent from client C to server S. According to Algorithm \ref{alg5}, if the push button is pressed, the client device sends the request message to a neighbor having $Min_{ID}$, that is, relay $a_1$ having ID 3. Therefore, client C sends the request message to relay $a_1$ instead of relay $a_{10}$. According to Algorithm \ref{alg4}, if the request message is received, the relay device sends it to  a neighbor having $Min_{ID}$. Therefore, relay  $a_{1}$ sends the request message to relay $a_{2}$. Afterwards, relay $a_{2}$ sends it to relay $a_{3}$ having ID 1. Finally, relay $a_{3}$ sends the request message to server S having ID 0. 

Figure 7c shows how the Ack message is sent back to the client device from server S. According to Algorithm \ref{alg1}, server S sends back the acknowledgment message to relay $a_{3}$.  As described in Algorithm~ \ref{alg4}, if the acknowledgment message is received, the relay device sends the received acknowledgment message to  a neighbor having $Max_{ID}$. Therefore, relay  $a_{3}$ sends the acknowledgment message to relay  $a_{2}$ having ID 2. Afterwards, relay $a_{2}$ sends it to relay $a_{1}$ having ID 3. Finally, relay $a_{1} $ sends it to client~ C.

  \section{Performance Evaluation}
In this section, the performance of the RDSP scheme is compared with the UF scheme \cite{b30} by deploying the  real time multihop communication network. 
\subsection{Deployment Environment}
\begin{figure}[H]
  
  \centering
    \includegraphics[width=0.5\textwidth]{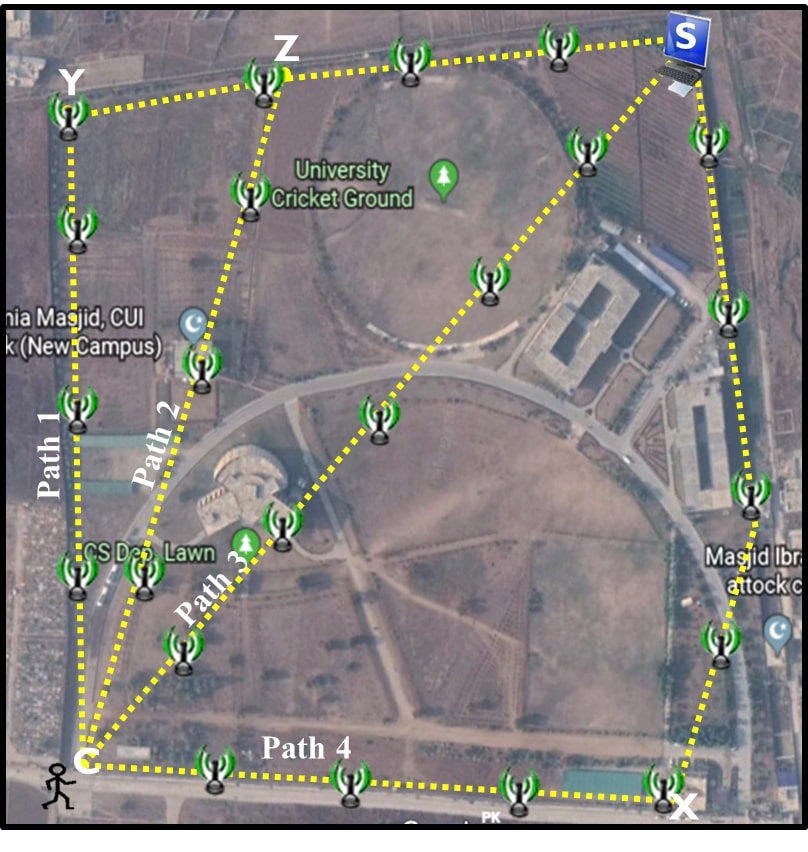}

     \caption{Deployment of the RDSP system.}
       \label{fig8}
  \end{figure}

A multihop communication network is deployed in Comsats university Islamabad-attock campus as shown in Figure \ref{fig8}. 

The total area of the university is 153,000  m². As shown in the figure, the server device is installed at point S and the client device is installed at point C. The relays are deployed all over the area in a manner to establish four different paths between server device S and client device C. The first path defined by points C, Y, Z, S contains 7 relays, second path defined by points C, Z, S contains 6 relays, third path defined by points C, S contains 5 relays and fourth path defined by points C, X, S contains 8 relays. The relays are placed at fixed positions at equal distance of 90 meters. To find the shortest paths between the client and server devices, the proposed RDSP system utilizes DIA and MMN algorithms whereas the UF scheme uses Destination-Sequenced Distance Vector (DSDV) \cite{b29} routing algorithm. In the DSDV, each node maintains a routing table with routes to destinations. An entry for a given destination consists of the ID of the next-hop, total hops to the destination, route's sequence number, and the time for the recent route update. In the case of the proposed RDSP, each node periodically broadcasts a hello message every 2 seconds containing only the node ID. However, in the UF scheme, each node periodically broadcasts an entire routing table every 1 second containing the route destination, the advertising node’s next-hop node for that route, the number of hops to the destination, and the sequence number. Following the deployment of the multihop communication system, both the client device and server device are tested to communicate wirelessly with each other across all four paths. One hundred request messages were generated to collect the results for each path. The experiment duration was set at 550 seconds. The results presented for each path are averaged over 5 repeated experiments. The experimental parameters are summarized in Table 1.


\begin{table}[H]
\caption{Experimental Parameters}
\centering
\begin{tabular}{ll}
\toprule
\textbf{Parameter}	& \textbf{Value}	\\
\midrule
Deployment scenario		& COMSATS University Campus			\\
Deployment area		& 153,000 m²			\\
Total relays	& 23			\\
Total Paths	& 4			\\
Relays Transmission range		& 90 meter			\\
Routing protocol (RDSP)  & DIA and MMN\\
Routing protocol (UF) & DSDV\\
Hello interval (RDSP)  & 2 sec\\
Hello inerval (UF) & 1 sec\\
Request messages & 100\\
Experiment duration & 550 seconds\\
\bottomrule
\end{tabular}
\end{table}


\subsection{Performance Metrics }
To investigate the performance of the RDSP scheme, the following metrics were used: 

\begin{itemize}
\item Distance covered: refers to the total distance covered by intermediate relays using wireless transmission
\item End-to-end delay: refers to the average time it takes for a request message sent from the client device to reach the server device.
\item Round-trip delay: refers to the average time it takes for a message sent from the client device to reach the server device and an acknowledgment message sent back from the server device to reach the client device.
\item Message delivery ratio: refers to the percentage of messages successfully received by the server device.
\item Network overhead: refers to the total number of control messages sent by relays on different paths.

\end{itemize}

\subsection{Distance Covered }
\begin{figure}[H]

  \begin{subfigure}[t]{0.9\textwidth}
  \centering
      \includegraphics[width=0.7\textwidth]{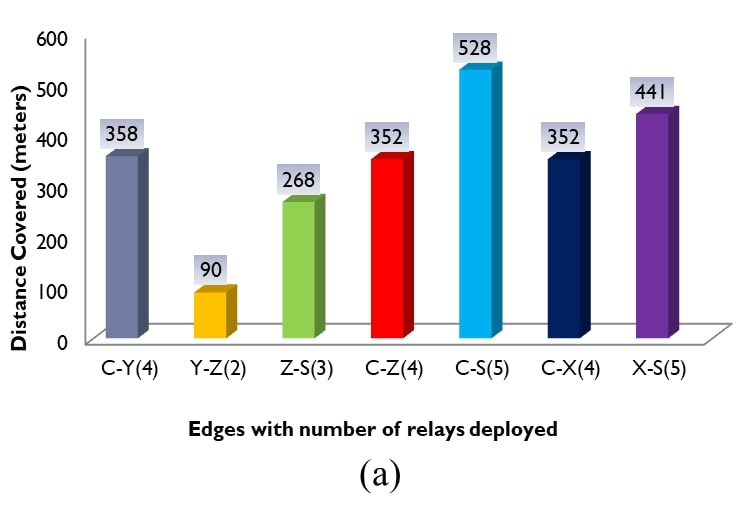}
     \end{subfigure}
   
     \begin{subfigure}[t]{0.9\textwidth}
     \centering
      \includegraphics[width=0.7\textwidth]{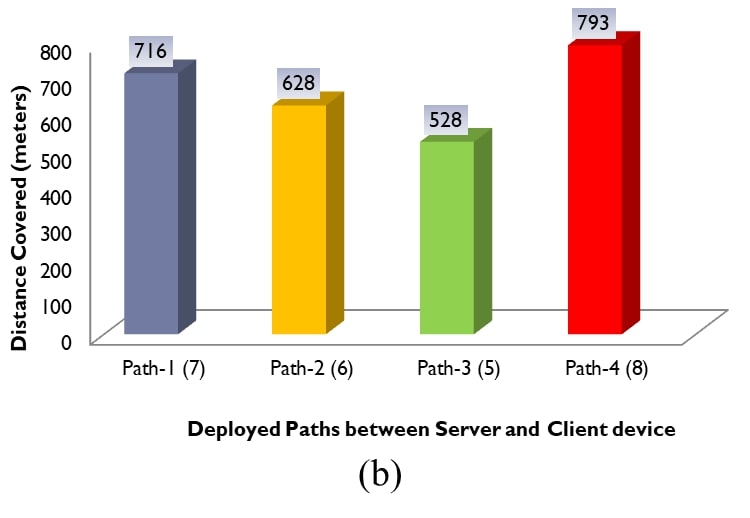}
     \end{subfigure}

    \centering
     \caption{Distance covered by edges and paths.}
       \label{fig9}
  \end{figure}

Figure \ref{fig9}a shows the distance covered by different edges described in Figure \ref{fig8} along with the deployed number of relays at each edge. Edge C-Y covers 358 meters with 4 relays whereas edge Y-Z and Z-S cover 90 and 268 meters with 2 and 3 relays respectively. Similarly, edge C-Z and C-S cover 352 and 528 meters with 4 and 5 relays, respectively. Finally, edge C-X and X-S cover 352 and 441 meters with 4 and 5 relays, respectively. It is evident from Figure \ref{fig9}a that as the number of relays increases across an edge, the distance covered also increases and vice versa.
Figure \ref{fig9}b shows the distance covered by four different paths described in Figure \ref{fig8} along with the number of relays deployed on each path. Distance covered by path-1 is 716 meters and comprises of three edges, i.e., C-Y, Y-Z and Z-S. Similarly, the distance covered by path-2 is 628 meters and comprises two edges, i.e., C-Z and Z-S. Likewise, path-3 covers 582 meters and comprises of edge C-S whereas path-4 covers 793 meters and comprises of edges C-X and X-S. It is once again obvious from Figure \ref{fig9}b that the paths comprising of more relays cover long distances compared with the paths comprising of less relays.

\subsection{ End-to-End Delay }

\begin{figure}[H]
\centering
  \begin{subfigure}[t]{0.8\textwidth}
      \includegraphics[width=0.9\textwidth]{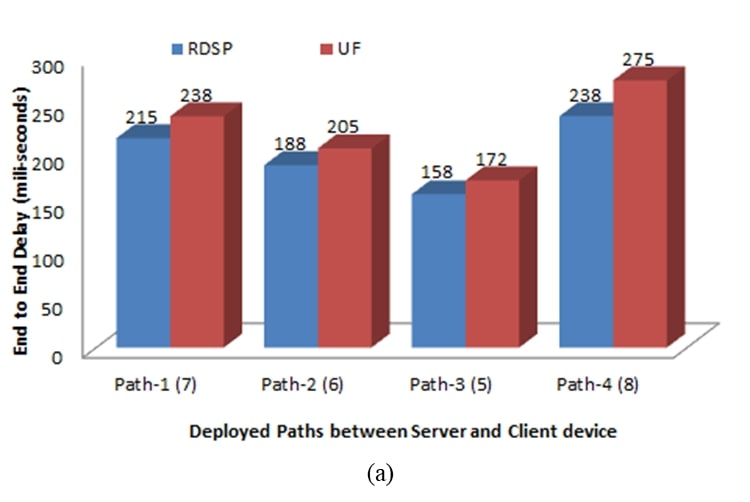}
     \end{subfigure}
   
     \begin{subfigure}[t]{0.8\textwidth}
      \includegraphics[width=0.9\textwidth]{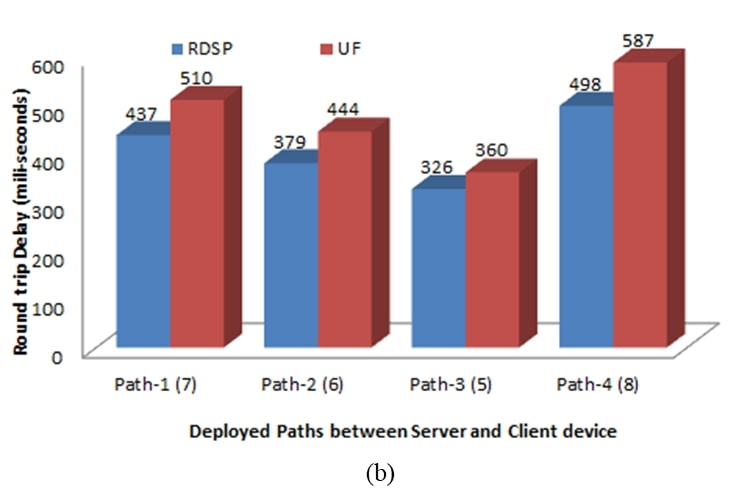}
     \end{subfigure}

     \caption{ End-to-end and round-trip delays.}
       \label{fig10}
  \end{figure}

Figure \ref{fig10}a compares the end-to-end delay for the RDSP system and UF scheme. It shows that the end-to-end delay for both systems increases as the number of relays increases across the path. To elaborate, the end-to-end delay for path-4 having 8 relays is greater than that of path-1 having 7 relays. Similarly, the end-to-end delay of path-1 is greater than that of path-2 having 6 relays. Likewise, the end-to-end delay of path-2 is greater than that of path-3 having 5 relays. This is because when the number of relays increases across a particular path, the processing and transmission time of messages also increase, but more relays offer a long-distance coverage benefit. However, the RDSP system achieved around 12\% lower end-to-end delays than the UF scheme across all the four paths. This was because the UF scheme is based on a DSDV protocol that requires all the relays to periodically exchange hello messages and entire routing tables, which leads to frequent contention and collisions among neighboring relays. In such cases, the relays must wait for a busy channel to become idle before performing any transmission. On the other hand, the RDSP system avoids the formation and exchange of routing tables and creates routes between the server and client device on the fly during the deployment of relays, as explained earlier in Figure \ref{fig7}, which eventually reduces collisions among neighboring relays and causes a decrease in the end-to-end delay. Figure \ref{fig10}b compares the round-trip delay for the RDSP system and UF scheme. It shows that the round-trip delays are approximately twice the end-to-end delays of both systems and R the DSP system achieves around 13\% lower round-trip delays than the UF scheme across all the four paths. 

\subsection{ Message Delivery Ratio}
\begin{figure}[H]
 
  \centering
    \includegraphics[width=0.8\textwidth]{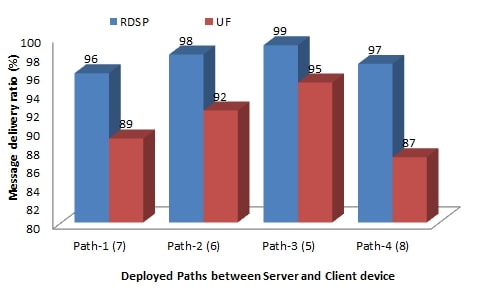}

     \caption{Message delivery ratio.}
       \label{fig11}
  \end{figure}

Figure \ref{fig11} compares the message delivery ratio for the RDSP system and UF scheme. It shows that the message delivery ration for both systems decreases as the number of relays increased across the path. To elaborate, the packet delivery ratio for path-4 having 8 relays is less than that of path-1 having 7 relays. Similarly, the packet delivery ration of path-1 is less than that of path-2 having 6 relays. Likewise, the packet delivery ratio of path-2 is less than that of path-3 having 5 relays. This was because when the number of relays increases across a particular path, the frequent contention and collisions of messages among neighboring relays also increases which eventually reduces packet delivery ratio. However, RDSP system achieved around 8\% higher message delivery ratio than UF scheme across all the four paths particularly at high relay densities because it avoids exchange of routing tables thus reduces collisions among neighboring relays and hence achieves higher message delivery ratio. On the other hand, UF scheme utilizes periodic exchange of entire routing tables that leads to frequent contention and collisions among neighboring relays which eventually reduces message delivery ratio.
\subsection{ Network Overhead}
\begin{figure}[H]
  
  \centering
    \includegraphics[width=0.8\textwidth]{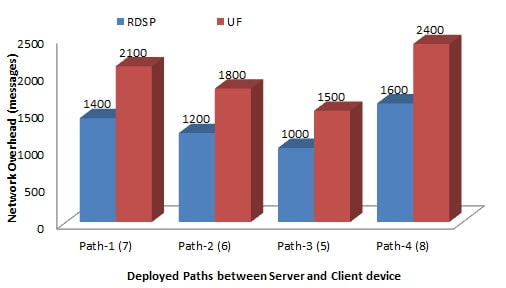}

     \caption{Network overhead.}
       \label{fig12}
  \end{figure}

Figure \ref{fig12} compares the average network overhead for the RDSP system and UF scheme. The UF overhead was very high because of the periodic exchange of entire routing tables to maintain the whole network information at each relay node. In contrast, the RDSP system overhead was much lower, due to the absence of the exchange of routing tables. Thus, the overhead in the RDSP system was related to the transmission of hello messages and acknowledgment messages. As a result, the average network overhead for the RDSP was about 33\% lower than that for the UF scheme.

\section{Conclusion}
In this study, an RDSP was proposed that aims to reduce the average waiting times for transmitting the request messages containing rescue groups and victim’s location information towards the control server. Additionally, unlike existing schemes, the RDSP does not rely on the periodic exchange of entire routing tables. However, the proposed RDSP scheme enables intermediate relays to dynamically select their IDs based on the information provided by their neighbor relays and then each intermediate relay selects the best forwarders towards the control server to minimize end-to-end delays. The results of real time experiments demonstrate that with the proposed RDSP scheme, a request message generated by the client device can reach the server device with a minimal delay in both light and heavily deployed paths compared to the existing UF system. In addition, the results confirmed that the RDSP outperforms the UF scheme under various relay densities in terms of the end-to-end delay, round-trip delay, massage delivery ratio, and network overhead.

\authorcontributions{ conceptualization, hardware, software, writing, A.K. and A.M.; methodology, validation, Z.K. and F.U.; resources, conceptualization, funding acquisition, M.B. and L.N.;  writing--original draft preparation, S.S.; resources, writing-review, and editing, L.D.N.; writing-review and editing, funding acquisition, S.M.R.I. and K.-S.K.} 

\textbf{funding}: This work was supported by National Research Foundation of Korea-Grant funded by the Korean Government (Ministry of Science and ICT) NRF 2020R1A2B5B0200247

\textbf{conflicts of interest}: {The authors declare no conflict of interest.}


\begin{thebibliography}{999}
\bibitem{b1} ITU. Measuring the Information Society Report 2018.


\bibitem{b2} Cangialosi, J.P.; Kimberlain, T.B. {\em Tropical Cyclone Report};  National Hurricane Center: Miami, FL, USA, September 2014. 


\bibitem{b3}  Mase, K. How to Deliver Your Message from/to a Disaster Area. \emph{IEEE Commun. Mag.} {\bf 2011}, {\em 49}, 52--57.



\bibitem{b4} Miranda, K.; Molinaro, A.; Razafindralambo, T. A survey on rapidly deployable solutions for post-disaster networks. \emph{IEEE Commun. Mag}. {\bf 2016}, {\em 54}, 117--123.



\bibitem{b5} Uddin, M.; Ahmadi, H.; Abdelzaher, T.; Kravets, R. A low-energy, multi-copy inter-contact routing protocol for disaster response networks. In Proceedings of the 6th Annual IEEE Communications Society Conference on Sensor,Mesh and Ad Hoc Communications and Networks, Rome, Italy, 22--26 June 2009; pp. 1--9.



\bibitem{b6} Guo, W.; Huang, X. Mobility model and relay management for disaster area wireless networks. In Proceedings of the Third International Conference (WASA 2008), Dallas, TX, USA, 26--28 October 2008; pp. 274--285.



\bibitem{b7} Jahir, Y.; Atiquzzaman, M.;  Refai, H.;  Paranjothi, A.; LoPresti, P. G.  Routing protocols and architecture for disaster area network: A survey. \emph{Ad Hoc Netw.} {\bf 2019}, {\em 82}, 1--14.



\bibitem{b8} Reina, D.; Askalani, M.; Toral, S.; Barrero, F.; Asimakopoulou, E.; Bessis, N. A Survey on Multihop Ad Hoc Networks for Disaster Response Scenarios. \emph{Int. J. Distrib. Sens. Netw.} {\bf 2015}, {\em 11}, 647037. 


\bibitem{b9} Li, H.; Shan, L.; Matsuda, T.; Miura, R. Design and deployment of infrastructure-independent D2D networks without centralized coordination. In Proceedings of the International Symposium on Wireless Communication Systems,Brussels, Belgium, 25-28 August 2015; pp. 376-380.



\bibitem{b10} Ghaznavi, I.; Heimerl, K.; Muneer, U.; Hamid, A.; Ali, K.; Parikh, T.S.; Saif, U. Rescue base station. In Proceedings of the Fifth ACM Symposium on Computing for Development, San Jose, CA, USA, 5--6 December 2014; doi:10.1155/2012/614532





\bibitem{b11} Pezeshkian, N.; Nguyen, H.G.; Burmeister, A.  Unmanned Ground Vehicle Radio Relay Deployment System for Non-Line-of-Sight Operations. In Proceedings of 13th IASTED International Conference on Robotics and Applications, Würzburg, Germany, 29--31 August 2007; pp. 501–506; doi: 10.21236/ada475525





\bibitem{b12} Nguyen, C.Q.; Min, B.-C.; Matson, E.T.; Smith, A.H.; Dietz, J.E.; Kim, D. Using mobile robots to establish mobile wireless mesh networks and increase network throughput. \emph {Int. J. Distrib. Sens. Netw.} {\bf 2012}, {\em 8}, 1--13. 

\bibitem{b13} Jia, S.; Fadlullah, Z.M.; Kato, N.; Zhang, L. Eco-Udc: An energy efficient data collection method for disaster area networks. In Proceedings of 2016 IEEE International Conference on Network Infrastructure and Digital Content (IC-NIDC), Beijing, China, 23--25 September 2016; pp. 130–134.
\bibitem{b14} Li, M.; Nishiyama, H.; Kato, N.; Owada, Y.; Hamaguchi, K. On the energy-efficient of hroughput-based scheme using renewable energy for wireless mesh net- works in disaster area. \emph{IEEE Trans. Emerg. Top. Comput.} {\bf{2015}}, \emph {3}, 420–431.
\bibitem{b16} Suzuki, H.; Kaneko, Y.; Mase, K.; Yamazaki, S.; Makino, H. An ad hoc network in the sky, skymesh, for large-scale disaster recovery. In Proceedings of IEEE Vehicular Technology Conference, Melbourne, Australia, 7--10 May 2006; pp. 1--5.
\bibitem{b17} Sakano, T.; Kotabe, S.; Komukai, T.; Kumagai, T.; Shimizu, Y.; Takahara, A.; Ngo, T.; Fadlullah, Z.M.; Nishiyama, H.; Kato, N. Bringing movable and deployable networks to disaster areas: Development and field test of mdru.  \emph{IEEE Netw.} {\bf{2016}}, \emph{30}, 86–91. 
\bibitem{b18} Ngo, T.; Nishiyama, H.; Kato, N.; Sakano, T.; Takahara, A. A spectrum-and energy–efficient scheme for improving the utilization of MDRU-based disaster resilient networks. \emph{IEEE Trans. Veh. Technol.} {\bf{2014}}, \emph{63}, 2027--2037.
\textcolor{black}{
\bibitem{b19}	Chen, A.Y.; Peña-Mora, F.; Plans, A.P.; Mehta, S.J.; Aziz, Z. Supporting Urban Search and Rescue with digital assessments of structures and requests of response resources. \emph{Adv. Eng. Inf.}  {\bf{2012}}, \emph {26}, 833--845.}


\bibitem{b20} Hu, D.; Li, S.; Chen, J.; Kamat, V.R. Detecting, locating, and characterizing voids in disaster rubble for search and rescue. \emph{Adv. Eng. Inf.}  {\bf{2019}}, \emph {42}, 100974.
\textcolor{black}{
\bibitem{b21}	Peña-Mora, F.; Chen, A.Y.; Aziz, Z.; Soibelman, L.; Liu, L.Y.; El-Rayes, K.; Arboleda, C.A.; Lantz, Jr. T.S.; Plans, A.P.; Lakhera, S.; et al. Mobile Ad Hoc Network-Enabled Collaboration Framework Supporting Civil Engineering Emergency Response Operations.\emph { J. Comput. Civil Eng.}  {\bf 2010}, \emph {3}, 302--312}.
\textcolor{black}{
\bibitem{b22}Chen, A.Y.; Peña-Mora, F.; Ouyang, Y. A collaborative GIS framework to support equipment distribution for civil engineering disaster response operations. \emph {Autom. Constr.}  {\bf{2011}}, \emph {20}, 637--648.}
\textcolor{black}{
\bibitem{b23}	Crawford, P. S.; Al-Zarrad, M. A.; Graettinger, A. J.; Hainen, A. M.; Back, E.; Powell, L.  Rapid Disaster Data Dissemination and Vulnerability Assessment through Synthesis of a Web-Based Extreme Event Viewer and Deep Learning. \emph {Adv. Civ. Eng.} {\bf 2018}, {\em 2018}; doi:10.1155/2018/7258156.}
\textcolor{black}{
\bibitem{b24}	Menon, V.G.; Pathrose, J.P.; Priya, J. Ensuring Reliable Communication in Disaster Recovery Operations with Reliable Routing Technique. \emph{Mob. Inf. Syst.} {\bf 2016}, {\em 2016}; doi:10.1155/2016/9141329.}
\textcolor{black}{
\bibitem{b25}	Vo, N.-S.; Masaracchia, A.; Nguyen, L.D.; Huynh, B.-C. Natural Disaster and Environmental Monitoring System for Smart Cities: Design and Installation Insights. \emph{ EAI Endorsed Trans. Indust. Netw. Intellig. Syst.} {\bf{2018}}, {\em 5}, e5.}
\textcolor{black}{
\bibitem{b26}	Nguyen, L.D.; Kortun, A.; Duong, T.Q. An Introduction of Real-Time Embedded Optimisation Programming for UAV Systems under Disaster Communication.  \emph{ EAI Endorsed Trans. Indust. Netw. Intellig. Syst.} {\bf{2018}}, \emph {5}, e5.}
\bibitem{b27} Souryal, M.R.; Geissbuehler, J.; Miller, L.E.; Moayeri, N. Real-Time Deployment of Multihop Relays for Range Extension. In Proceedings of the 5th international conference on Mobile systems, applications and services. (MobiSys), San Juan, PR, USA, 11--14 June 2007; pp. 85--98, doi:10.1145/1247660.1247673.

\bibitem{b28} 	Wolff, A.; Subik, S.; Wietfeld, C. Performance analysis of highly available ad hoc surveillance networks based on dropped units. In Proceedings of the 2008 IEEE Conference on Technologies for Homeland Security, Waltham, MA, USA, 12--13 May 2008; pp. 123--128.


\bibitem{b29} Bao, J.Q.;  Lee, W.C. Rapid deployment of wireless ad hoc backbone net works for public safety incident management. In Proceedings of the  IEEE GLOBECOM 2007---IEEE Global Telecommunications Conference, Washington, DC, USA, 26--30 November 2007; pp.  1217--1221.

\bibitem{b30} 	Liu, H.; Xie, Z.; Li, J.; Li, S.; Siu, D.J.; Hui, P.; Whitehouse, K.; Stankovic, J.A. An automatic, robust, and efficient multi-user breadcrumb system for emergency response applications. \emph{IEEE Trans. Mob. Comput.} {\bf 2013}, \emph {13}, 723--736.


\bibitem{b31} Khan, A.; Ullah, F.; Kaleem, Z.; Rehman, S.U.; Anwar, H.; Cho, Y.-Z. EVP-STC: Emergency Vehicle Priority and Self-Organising Traffic Control at Intersections Using Internet-of-Things Platform.  \emph{IEEE Access} {\bf 2018}, \emph {6}, 68242--68254.
\end{thebibliography}
\end{document}